\documentclass[preprint]{aastex}

\begin{document}
 \pagestyle{plain}
 
 \title{Infrared Monitoring of the Microquasar GRS 1915+105: Detection of Orbital and Superhump Signatures}
 \author{Ethan T. Neil, Charles D. Bailyn, and Bethany E. Cobb}
 \affil{Department of Astronomy, Yale University}
 \affil{New Haven, CT}
 \email{ethan.neil@yale.edu, bailyn@astro.yale.edu, cobb@astro.yale.edu}

 \begin{abstract}
 We present the results of 7 years of $K$-band monitoring of the low-mass X-ray binary GRS 1915+105.  Positive correlations between the infrared flux and the X-ray flux and X-ray hardness are demonstrated.  Analysis of the frequency spectrum shows that the orbital period of the system is $P_{orb}= 30.8 \pm 0.2$ days.  The phase and amplitude of the orbital modulation suggests that the modulation is due to the heating of the face of the secondary star.  We also report another periodic signature between 31.2 and 31.6 days, most likely due to a superhump resonance.  From the superhump period we then obtain a range for the mass ratio of the system, $0.05 < q < 0.12$.
\end{abstract}
 
 \keywords{accretion, accretion disks --- black hole physics --- infrared: stars --- X-rays: binaries --- X-rays: individual (GRS 1915+105)}
 
\section{Introduction}\label{SecInt}
GRS 1915+105 is a well-known and somewhat unusual example of a galactic X-ray binary system.  It was first observed in outburst on 1992 August 15 \citep{Castro-Tirado:1994ys}.  It is believed to contain a black hole that is both maximally rotating and rather massive at $14 \pm 4$ solar masses \citep{Greiner:2001kx}.  GRS 1915+105 was also the first known galactic example of a ``superluminal source" \citep{Mirabel:1994uq}, so called because the relativistic jets that are ejected from it move with an apparent velocity greater than $c$.  Unlike most transient low-mass X-ray binaries, GRS 1915+105 has never been observed in quiescence since it first became active.

High absorption along the line of sight to GRS 1915+105 makes observation in optical and IR wavelengths difficult.  However, a $K$-band counterpart has been discovered \citep{Mirabel:1994fk}.  Here we report daily monitoring of the $K$-band counterpart of GRS 1915+105.  In \S \ref{SecObs} we review our observation and data reduction methods, and present $K$-band and X-ray light curves.  Section \ref{SecCor} features analysis of the correlation between IR flux, X-ray flux, and X-ray hardness ratio.  In \S \ref{SecPer} we discuss extraction of the orbital period, its phase and amplitude, and physical implications.  Finally, \S \ref{SecSup} concerns the second, ``superhump" period, leading to limits on the mass ratio of the system.  Section \ref{SecCon} reiterates our conclusions.

\section{Observations and Data Reduction}\label{SecObs}
Our observations of GRS 1915+105 span seven seasons, from MJD 51652 to 53969 (2000 April 18 to 2006 August 21), with a total of 604 individual data points (see Table \ref{datatable}.)  Data from 2000-2002 were taken using the ANDICAM instrument\footnote{http://www.astro.yale.edu/smarts/ANDICAM/} and the Yale 1.0m telescope at the Cerro Tololo Inter-American Observatory (CTIO), which was operated at the time by the YALO consortium \citep{Bailyn:1999bh}.  Subsequent data were obtained using the same instrument mounted on the 1.3m telescope originally constructed for the Two Micron All Sky Survey (2MASS), also at CTIO and operated by the Small and Moderate Aperture Research Telescope System (SMARTS) consortium\footnote{http://www.astro.yale.edu/smarts/}.  All observations are internally dithered by a tiltable mirror to remove artificial effects such as bad pixels and to allow sky subtraction.  Due to the large extinction, our observations are in the $K$-band only.

$K$-band images of GRS 1915+105 were taken at five spatially offset positions, with five internally dithered images at each position for a total of 25 images per observation.  When combining the images the spatial offset was removed first, then the internal dithering; this process served to reduce artifacts introduced by irregularities in the filters during internal dithering.  Combination was done using the IRAF software package, along with sky and flat subtraction and bias removal.  Exposure time for each individual image was 60 s for data taken from 2000-2002, and 40 s for that from 2003-2006, for a total exposure time of 1500 s and 1000 s, respectively.  Due to the larger aperture of the 1.3m telescope, the signal-to-noise ratio generated by these exposure times is similar.

Differential photometry was performed on the reduced images in IRAF, by comparing the IR counterpart of GRS 1915+105 to a set of non-variable reference stars.  Three different sets of reference stars were used in all: one for data from 2000, one from 2001-2003, and one from 2004-2006.  Reference star sets were changed in the first instance to account for a bad quadrant on the detector, and in the second instance due to the change in the field of view as the instrument moved to a different telescope.  Calibrated magnitudes for the most recent data were obtained by comparison with 2MASS catalog values of several reference stars.  Earlier differential photometry done with different sets of reference stars was then standardized by subtracting the difference between the average instrumental magnitudes of each reference set.

X-ray counts are taken from the All-Sky Monitor (ASM) instrument aboard the \textit{Rossi X-ray Timing Explorer} (\textit{RXTE}) satellite,\footnote{http://xte.mit.edu/} over an energy range of 1.5-12 keV.  For the purpose of considering X-ray hardness we define ``hard" and ``soft" X-ray bands as those above and below 5 keV, respectively, corresponding to the ASM's own $C$ and $A+B$ bands.

\begin{figure}
   \centering
   \plotone{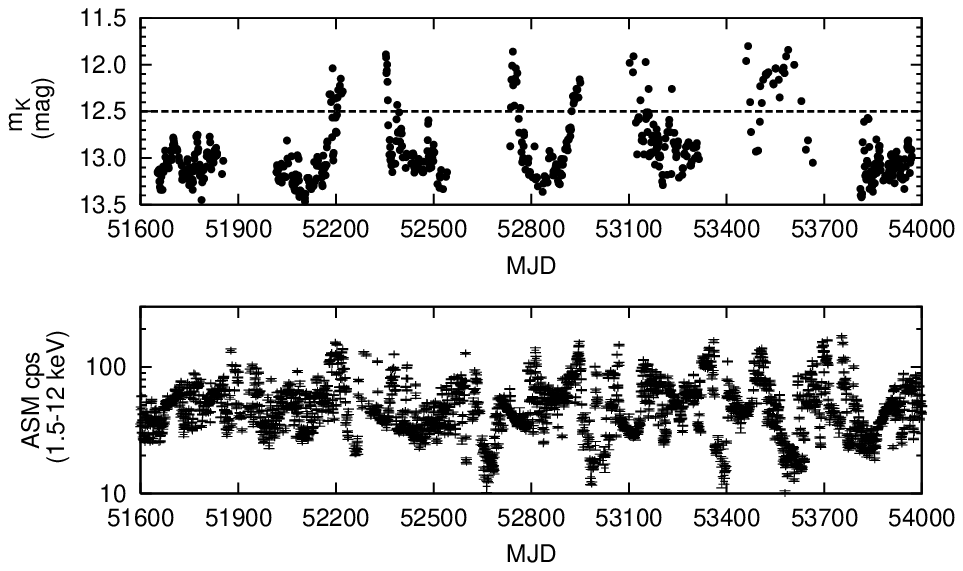}
   \caption{7-year accumulated light curve for GRS 1915+105, K-band (top panel) and X-ray (bottom panel).  The dotted line at $m_K = 12.5$ represents a magnitude cutoff used in the frequency analysis.}
   \label{LC}
\end{figure}

The combined $K$-band and X-ray light curve is presented in Figure \ref{LC}.  The large gaps seen in the $K$-band data correspond to seasons when the line of sight to GRS 1915+105 was near the Sun, making observation impossible.  The apparent rough correlation between the two bands, particularly near peak values, is characteristic of an X-ray binary system.  Note that the X-ray counts per second are always significant, so that the system never truly reaches a quiescent state over the course of our observations.

The dotted line at $m_K = 12.5$ indicates a magnitude cutoff used in the frequency analysis.  In performing a frequency analysis, we are looking in particular for luminosity variations related to the orbital period of the secondary star.  Regardless of the exact nature of such magnitude variations, we expect generically that they will be drowned out by emission from the accretion disc when the system is brightest.  Including data in which the period is not detectable will only reduce the signal-to-noise ratio, so we excluded all points above a fixed magnitude cutoff from the calculation of the frequency spectrum.  Interpretation of this method in light of our particular model of the system, as well as a justification for the choice of the value $12.5$ mag, are given in \S \ref{SecPer} below.

\begin{figure}
   \centering
   \epsscale{1.10}
   \plottwo{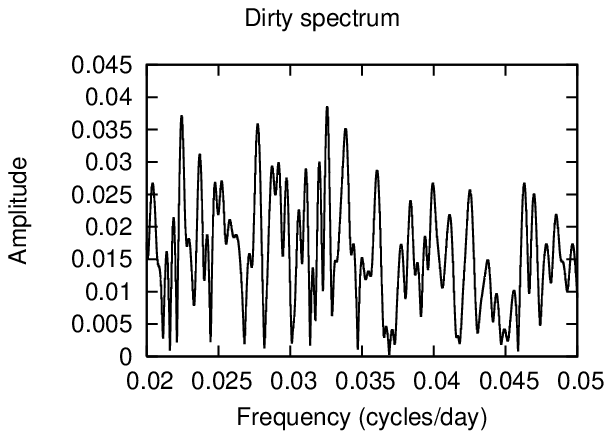}{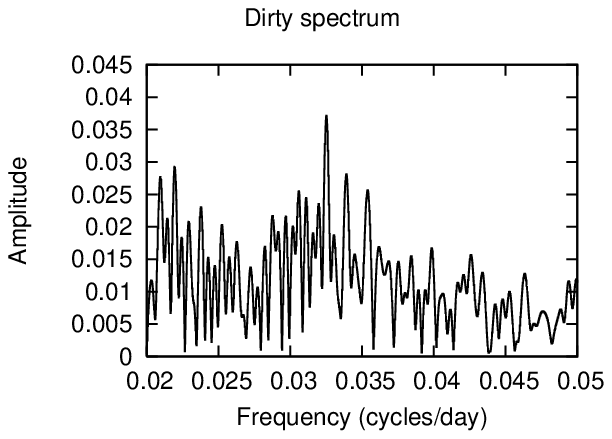}
   \plottwo{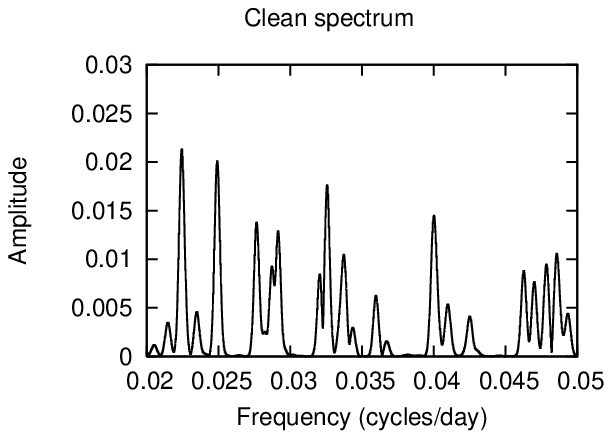}{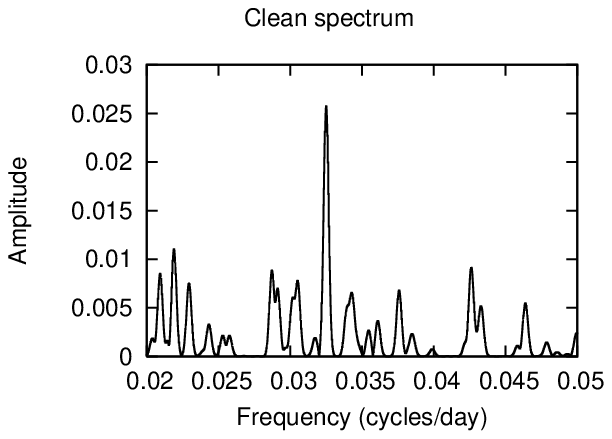}
   \caption{Visual demonstration of the effect of the CLEAN algorithm.  The left column is derived using the entire data set, while the right column excludes data with $m_K < 12.5$.  In each column, the top plot is the raw spectrum, calculated directly from the data; the bottom plot is the cleaned spectrum.  Note that the orbital signal near $0.033$ cycles/day is present in both plots, but stands out much more clearly when only the faint data are considered.}
   \label{clean}
\end{figure}

To search for periodic structure in the data, the raw $K$-band results were transformed to the frequency domain.  For time-domain data sets with large, irregular gaps such as ours, attempting a naive Fourier analysis would introduce a large number of artifacts into the frequency domain, obscuring the true signature from the data.  To avoid this effect, we used a one-dimensional implementation of the well-known CLEAN algorithm \citep{Roberts:1987fk} that reduces the number of artifacts introduced into the power spectrum.  The effect of ``CLEANing" on the raw, or ``dirty," spectrum is shown in Figure \ref{clean}, including a demonstration of the effect of the magnitude cut.

\section{X-ray/IR correlations}\label{SecCor}
Conversion of $K$-band apparent magnitudes into fluxes was accomplished with reference to \cite{Bessell:1988qf}, according to which the two quantities are related by $m = -2.5 \log_{10} (f_\nu) - 66.08 - \textrm{zp}(f_\nu)$.  Here zp$(f_\nu)$ is a zero-point, which is given to be $1.88$ for the $K$-band.  In the above formula, $f_\nu$ is taken to be in units of J s$^{-1}$ cm$^{-2}$ Hz$^{-1}$, which is equivalent to $10^{36} \mu$Jy.

\begin{figure}
   \centering
   \epsscale{1.00}
   \plotone{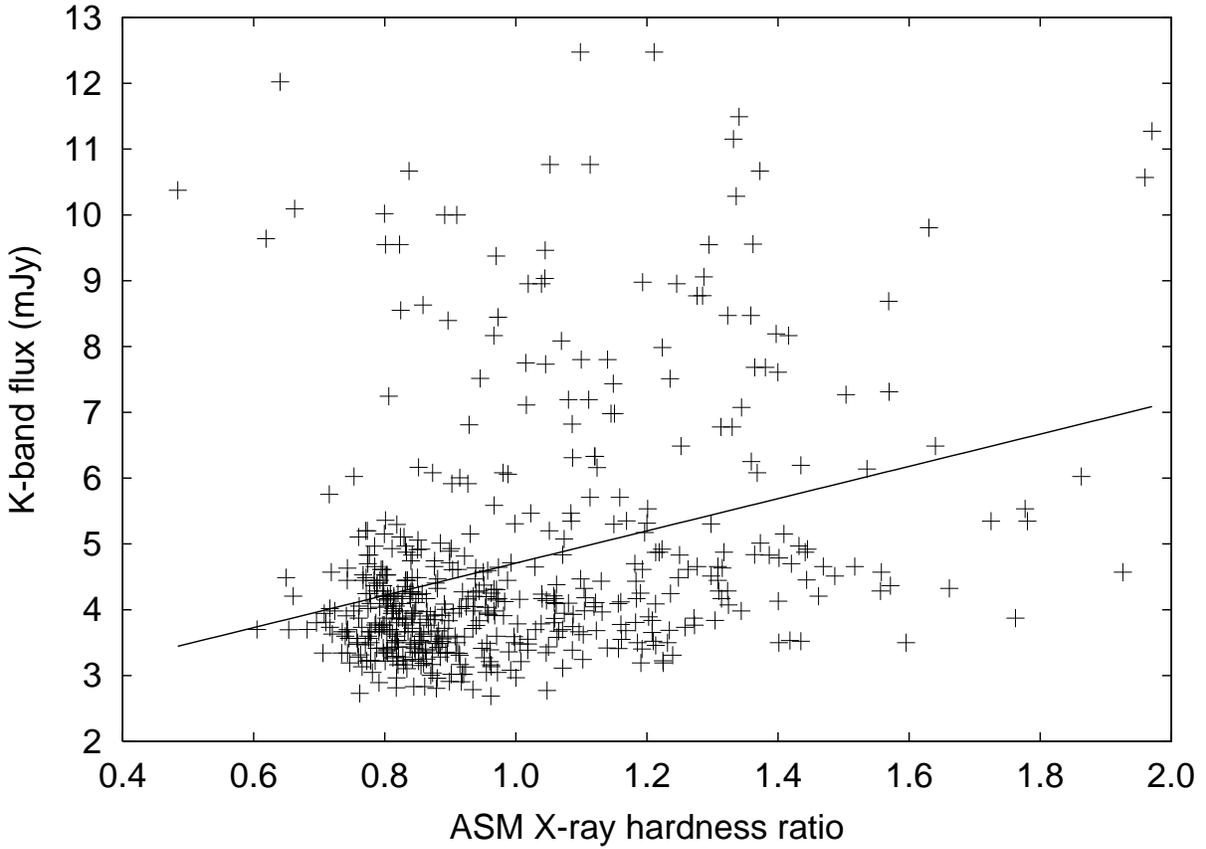}
   \caption{K-band flux versus X-ray hardness ratio.  Hardness ratio is simply the ratio of hard to soft x-rays, where ``hard" and ``soft" are defined as more or less energetic than 5 keV, respectively.  The line depicted is the least-squares best fit to the data.}
   \label{scatter}
\end{figure}

Least-squares linear fitting of our results demonstrates a correlation between the X-ray and IR data.  In particular both the X-ray counts per second and hardness ratio are positively correlated with the $K$-band flux; this result is consistent with previous analysis of GRS 1915+105 \citep{Greiner:2001uq}.  The correlation coefficients obtained are $r = 0.194$ and $0.330$, respectively, so the IR/hardness correlation is stronger.  Both correlation coefficients are found to be different from zero at a $p$-level below $10^{-6}$ \citep[p.200]{Bevington:1992uq}, indicating that our result is statistically significant.  The relation between $K$-band flux and X-ray hardness ratio is depicted in Figure \ref{scatter}.

It is clear from figure \ref{scatter} that there is no significant correlation at large $K$-band flux.  Since the correlation is established entirely by the points with low $K$ flux, our linear fit to the full data set is not physically relevant.  This suggests that there are several sources of IR flux, not all of which are correlated with the X-ray properties of the source.  Thermal IR flux from the outer parts of the accretion disk is expected to correlate with the X-rays, since both depend on the accretion flow through the disk.  On the other hand, large non-thermal IR flares without an easily observed X-ray response have been reported in several other transients \citep{Buxton:2004zr,Jain:2001ly}.  We therefore suggest that the IR flux observed in GRS 1915+105 includes both a thermal component associated with the disk, and occasionally larger, possibly non-thermal emission uncorrelated with the X-rays.  We also expect a contribution from the secondary star, which will be discussed in more detail in the next section.  Unfortunately, our ability to interpret the IR flux in greater detail is limited by the fact that we have IR observations in only one band, and so have no IR color information.

\section{Orbital period, phase, and amplitude}\label{SecPer}

\begin{figure}
   \centering
   \plotone{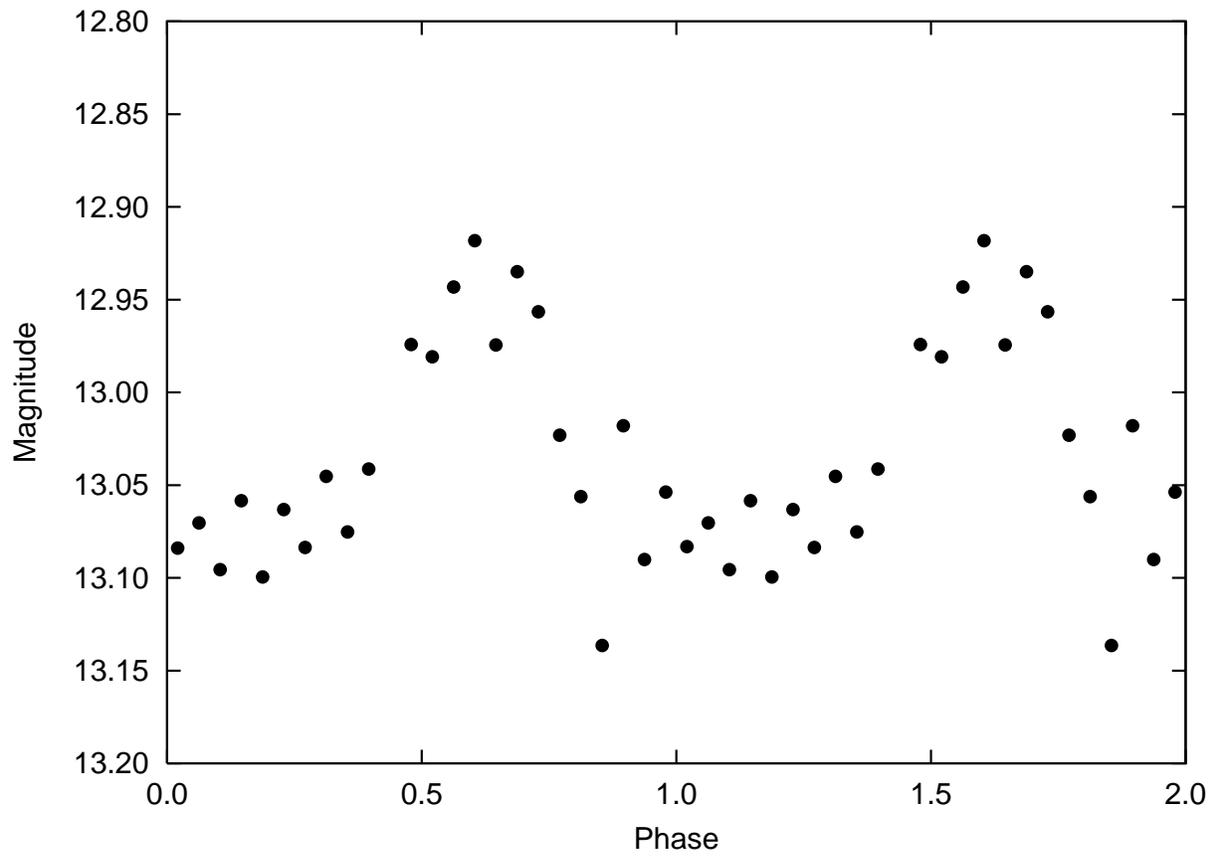}
   \caption{Folded phase plot of the K-band data below magnitude 12.5.  Data are binned and averaged according to phase, assuming a periodicity of $30.8$ days.  48 bins are used over a single period.}
   \label{foldphase}
\end{figure}

Conversion of our IR observations into frequency space, CLEANed and cut at 12.5 mag, reveals a significant peak corresponding to a period of $P_{orb} = 30.8 \pm 0.2$ days.  The error is estimated by fitting a Gaussian to the frequency spectrum at the peak frequency and taking the standard deviation of the fit to be the associated error $\sigma$.  Monte Carlo statistical testing \citep{Nemec:1985fk} over $5000$ trials yields a significance level of the period that is indistinguishable from $1$.  Folding and binning the data according to phase at this period makes the periodic structure manifest in the time domain (see Figure \ref{foldphase}.)  Furthermore, our period agrees to $< 2 \sigma$ with previous spectroscopic observations of GRS 1915+105, which determined an observed period of $33.5 \pm 1.5$ days \citep{Greiner:2001kx}.

For $P_{orb} = 30.8$ days, we find that our phase corresponds to a point of minimum brightness at MJD $53945.7 \pm 0.2$.  Early in our light curve, minimum brightness occurred at MJD $51666.5 \pm 0.2$, in extremely good agreement with the blue-to-red radial velocity crossing observed by \cite{Greiner:2001kx} at MJD $51666 \pm 1.5$.  This agreement is consistent with interpretation of the orbital modulation as being due to the heated face of the secondary star.

Before proceeding with our analysis of the frequency spectrum, we introduce a simple physical model of the system, assuming that the heated face of the secondary is responsible for the orbital variation.  In addition to lending physical significance to our measurements, the model allows us to validate our method of omitting data below a particular cutoff from our analysis.  In particular we seek to describe the  size of the variation of the observed magnitude due to the orbital period of the donor star.

If the side of the donor star facing the compact object is heated by emission from the accretion flow, then brightness variations will occur at the orbital period due to the rotation of the star, which is tidally locked to the orbital period.  Although the detailed behavior is complex, this model features a simple and robust prediction of the phase dependence.  By purely geometric reasoning, we can see that the phase corresponding to minimum measured brightness also corresponds to the point of blue-to-red crossing in the companion star's radial velocity, when the heated face is completely occluded.  As noted above, the observed phase of our photometric observations agrees with that of the spectroscopy of \cite{Greiner:2001kx}, confirming this prediction.

\begin{figure}
   \centering
   \epsscale{1.10}
   \plottwo{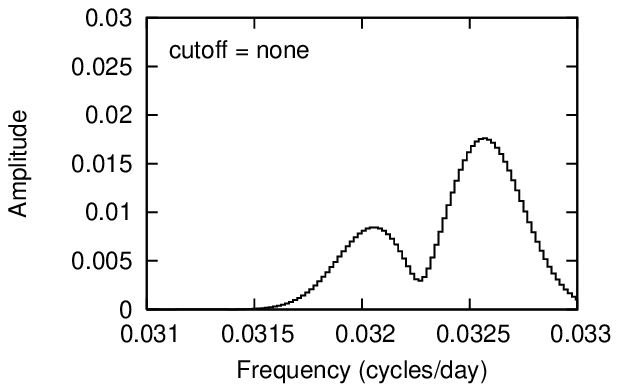}{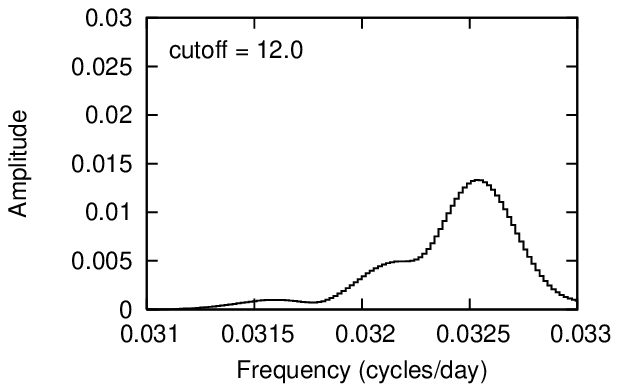}
   \plottwo{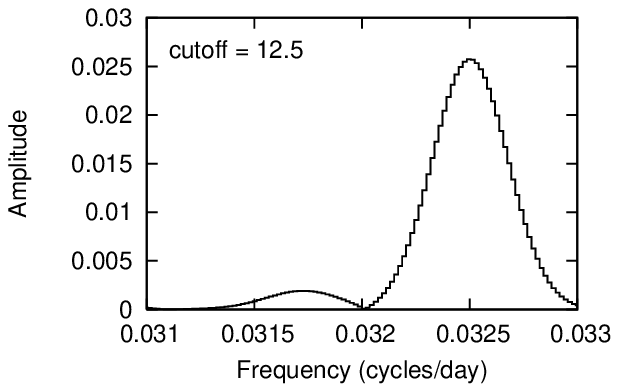}{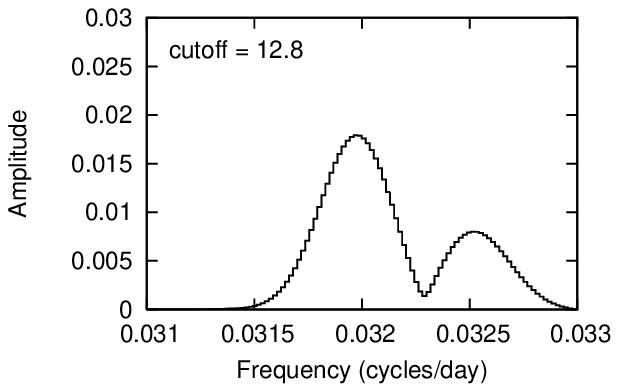}
   \caption{Effect of various magnitude cutoffs on the frequency spectrum around the peak of interest.  Clockwise from the top left, the depicted cuts are none, 12.0, 12.8, and 12.5 mag, where a cut at $m$ omits all magnitudes brighter than $m$.  The orbital frequency remains stable around 0.0325 $d^{-1}$, regardless of cut.}
   \label{cuts}
\end{figure}

Based on this model of GRS 1915+105, our use of a magnitude cutoff seems to be reasonable.  When the accretion disc is emitting strongly we would expect the brightness variations due to the heated face of the star to be diluted.  The exact choice of $m_K = 12.5$ for the cut is arbitrary, but not unreasonable.  
Through experimentation with various cuts, we determined that the detected period of interest is quite robust against the choice of cut (see Figure \ref{cuts}).  We adopted the cut at $12.5$ mag because it yields the strongest orbital signal.  Presumably, cuts below this value reduce the number of relevant data points included, while cuts above this value include points at which the orbital variation is suppressed by strong emission from the compact object; both result in a reduced signal-to-noise ratio.  The cut is depicted against the light curve in Figure \ref{LC}.

In order to interpret the amplitude of the Fourier component at $P_{orb}$, we must adopt some additional assumptions.  We model the heated face of the star as a sphere which has one hemisphere brighter than the other.  Adding a coordinate system and doing some simple vector analysis then yields an expression for $\Delta F_K$, the variable part of the flux from the system.  The resulting formula is

\begin{equation}
\Delta F_K (\theta) = \frac{k F_0}{2} (1 - \frac{2\theta}{\pi})
\end{equation}

where $F_0$ is the flux of the total system in the absence of heating, $k$ is a constant of proportionality relating the variation $\Delta F_K$ to the total emitted flux $F_0$, and $\theta$ is the angle between the line of sight and the line from the compact object to the star.  Note that the range of $\theta$ is determined by the inclination angle $i$ of the orbital system: $\theta \in [\pi/2 - i, \pi/2 + i]$.  In particular if $i = 0^\circ$ there is no variation in the flux, since at all times half of each of the bright and dark hemispheres will be visible.  

Converting to magnitudes and taking the total range of variation, we find
\begin{equation}
\Delta m_K = -2.5 \log \left( \frac{2-ki}{2+ki} \right)
\end{equation}
where $i$ should be expressed in radians.  At $P_{orb} = 30.8$ days, we find the amplitude of the magnitude variation (i.e. the magnitude of the CLEANed frequency spectrum component at the period of interest) to be $0.034$ mag; the total range of variation $\Delta m_K$ is equal to twice this.  Using an inclination angle of $i = 70^{\circ} \pm 2^{\circ}$ \citep{Greiner:2001kx}, we find $k = 0.051 \pm 0.001$.  In other words, our model predicts that the flux emitted by GRS 1915+105 is roughly $5\%$ greater when the heated face of the secondary star is maximally visible.

The magnitude variation can also be related to the temperature difference $\Delta T$ between the two hemispheres of the star.  We assume a blackbody spectrum, so that the distribution of intensity with respect to wavelength follows Planck's Law.  Integrating over the range of wavelengths corresponding to our $K$-band filter then yields the $K$-band flux.

By determination of the spectral type, the temperature of the secondary star has been estimated to be $T_0 = 4800$ K \citep{Greiner:2001vn}.  Our $K$-band filter includes wavelengths on the order of $2 \mu $m, which is well into the tail of the Planck distribution at the given temperature.  Therefore we take the $kT \gg 1$ approximation, which gives a flux directly proportional to temperature.  In terms of $\Delta m_K$, we find 

\begin{equation}
\Delta T = T_0 \left[ 1 - \left(10^{-\Delta m_K / 2.5} + \frac{F_c}{F_0} (10^{-\Delta m_K / 2.5} - 1)\right)\right]
\end{equation}
 
 where $T_0$ is the temperature of the unheated side of the star, and $F_c / F_0$ is the ratio between the flux emitted by the compact object and accretion flow ($F_c$) and the unheated side of the star ($F_0$).  Taking $\Delta m_K = 0.068$ as above and $T_0 = 4800$ K as noted, we find a lower bound of $\sim 600$ K on the temperature difference if $F_c > F_0$, which is expected since GRS 1915+105 has never reached a quiescent state.  If the quiescent magnitude of the system is measured, then $F_0$ will be known, and $\Delta T$ and $F_c$ will be able to be determined for typical outburst values.
 
 \section{Superhump period and mass ratio}\label{SecSup}
 
 Also apparent in our frequency analysis was the existence of a period slightly longer than the orbital period.  Exact determination of this period is not possible, as it wanders over a range of values depending on which subset of the data is examined, as seen in Figure \ref{cuts}.  Specifically, it appears to range from $31.2$ to $31.6$ days.
 
\begin{figure}
   \centering
   \epsscale{0.70}
   \plotone{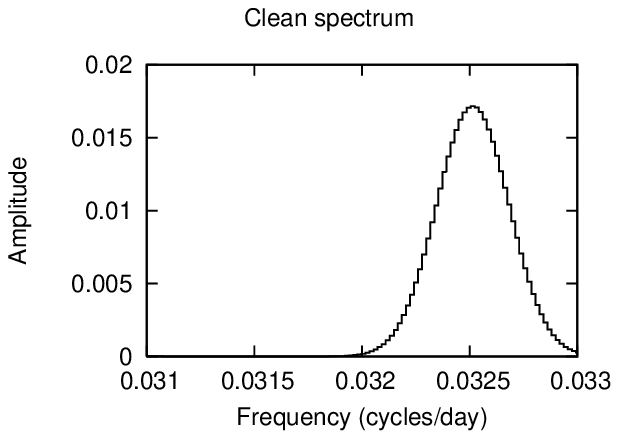}
   \plotone{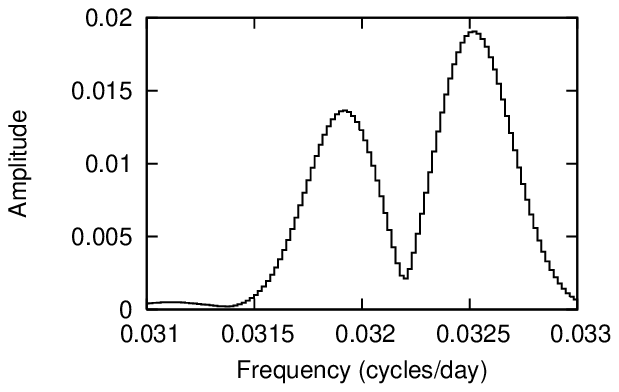}
   \caption{Clean spectra for different subsets of data.  For the top graph, only ``rising" data are kept, i.e. points at which the K-band magnitude is (locally) increasing.  The lower graph uses only ``falling" data.  Both plots also include a cutoff at $12.5$ mag.}
   \label{superhump}
\end{figure}

 Separating our data into ``rising" and ``falling" subsets, corresponding to how the magnitude is varying over sets of three adjacent data points, produces a striking difference in the frequency spectrum, depicted in Figure \ref{superhump}.  The secondary period is strongly visible in the falling data, but vanishes completely in the rising data.  The peak observed in the falling plot corresponds to a period of $31.4$ days.
 
 A period such as this one, differing by a few percent from the orbital period of the system, is indicative of a superhump, a resonance in the accretion disc itself frequently observed in cataclysmic variables (CVs; \cite{Whitehurst:1988dq,Patterson:2005qy}), as well as low-mass X-ray binaries \citep{ODonoghue:1996kx,Zurita:2002fj}.  The mass ratio of GRS 1915+105 is certainly well below the maximum value of $0.33$ for the existence of a superhump resonance \citep{Haswell:2001uq}.  The difference between the ``rising" and ``falling" data is also consistent with previous observations of superhumps, which are typically only seen in systems declining from outburst, although this may be somewhat of an empirical rule since observation of the rising magnitudes leading up to outburst is difficult in CVs.    Since $P_{sh}$ is strictly larger than $P_{orb}$, we can conclude that the superhump precession is prograde with respect to the rotation of the accretion disc \citep{Whitehurst:1991fk}.

 Finally, we determine the quantity $\epsilon$, the fractional difference between the superhump and orbital periods, which is defined as $P_{sh}/P_{orb} - 1$.  Given our range of values for $P_{sh}$, we find that $\epsilon$ lies between $0.010$ and $0.026$.  There is a simple empirical formula relating $\epsilon$ to the mass ratio of the system $q$, derived from observation of a number of low-mass X-ray binary and CV systems \citep{Patterson:2005qy}:
 
 \begin{equation}
 \epsilon = 0.18q + 0.29q^2.
 \end{equation}
 
 By this rule, the mass ratio $q$ of GRS 1915+105 should lie between $0.05$ and $0.12$.  This is in agreement with \cite{Greiner:2001kx}, who find a mass ratio of $q = 0.086 \pm 0.028$.  Their result depends on assigning a mass to the observed spectral type of the secondary star, rather than the more reliable line-broadening technique (e.g. \cite{Orosz:1994ve}).  Our result confirms that the masses deduced by Greiner et al. for the components of this system are approximately correct.  Conversely, if we assume $q = 0.086$ then $\epsilon(q) =  0.018$, corresponding very closely to the superhump peak seen in the ``falling" data subset (Figure \ref{superhump}.)
 
\section{Conclusion}\label{SecCon}
 
The most significant periodicity seen in the frequency spectrum of our data is a feature at a period of approximately $31$ days.  This is consistent with the spectroscopically determined orbital period of the secondary star, determined by \cite{Greiner:2001kx}, of $33.5 \pm 1.5$ days.  Therefore we claim $P_{orb} = 30.8 \pm 0.2$ days to be the orbital period, with a phase corresponding to minimum light at $T_0 =$ MJD $53945.7 \pm 0.2$.

In addition, we find in our data another period ranging from $31.2$ to $31.6$ days, which we attribute to a ``superhump" resonance in the accretion disc \citep{Haswell:2001uq}.  The superhump period is longer than the orbital period, indicating that the precession is prograde with respect to the rotation of the disc \citep{Whitehurst:1991fk}.  The fractional difference $\epsilon$ between the orbital and superhump periods is determined to lie between $0.010$ and $0.026$.  Our measurement of $\epsilon$ allows us to make a prediction of the mass ratio $q$ of the system, based on an empirical formula \citep{Patterson:2005qy}.  We find that $0.05 < q < 0.12$, providing an independent verification of the result $q = 0.09 \pm 0.03$ due to \cite{Greiner:2001kx}.
 
The verification of the mass ratio $q$ via the superhump period is of particular importance.  The previous determination of the mass ratio was based on spectroscopic identification of the secondary star \citep{Greiner:2001vn}, which requires a number of assumptions; our method is based on an unrelated set of assumptions.  Conversely, our result can be seen as a verification of the empirical relation between $\epsilon$ and $q$, in a system with an extremely long orbital period compared to other known superhumpers \citep{Patterson:2005qy}.
 
 \section*{Acknowledgements}\label{SecAck}
 We thank the ASM/\textit{RXTE} teams at MIT and the \textit{RXTE} Science Operations Facility and Guest Observer Facility at NASA's Goddard Space Flight Center for providing the X-ray results.  We are grateful for the efforts of YALO/SMARTS service observers D. Gonzalez and J. Espinoza, who collected the data reported here.  This work was supported by NSF grant AST 04-07063 to C.D.B.

\begin{deluxetable}{cc}
\tablewidth{0pt}
\tablecaption{Sample of full K-band data set.\label{datatable}}
\tablehead{\colhead{MJD} & \colhead{$m_K$ (mag)}}
\startdata
51652.3878 & 13.15 \\
51653.3471 & 13.16 \\
51654.3760 & 13.16 \\
51655.4107 & 13.13 \\
51657.3792 & 13.25 \\
51658.4115 & 13.12 \\
51659.3678 & 13.31 \\
51660.3831 & 13.25 \\
51661.3867 & 13.34 \\
51663.3795 & 13.20 \\
\enddata
\tablecomments{The full data table is available in the electronic edition of the Astrophysical Journal.}
\end{deluxetable}

 \end{document}